\documentclass[twocolumn,aps,prb,graphicx,showpacs]{revtex4}

\usepackage{graphicx}
\begin{document}

{\bf Comment on "The nature of slow dynamics in a minimal model of
frustration limited domains" by P. L. Geissler and D. R. Reichman:} In a
recent paper Geissler and Reichman\cite{Geissler} investigated a model with
the Hamiltonian 
\begin{equation}
H=\int d{\bf r}\left[ \frac{1}{2}\phi \left( \tau +k_{0}^{-2}\left( \nabla
^{2}+k_{0}^{2}\right) ^{2}\right) \phi +\frac{\lambda }{4!}\phi ^{4}\right] 
\label{Braz}
\end{equation}%
with scalar field $\phi \left( {\bf r}\right) $ in three dimensions using
Monte Carlo simulations.\cite{fn1} The parameters of the model are the
typical wave number, $k_{0}$, for modulations of $\phi $, the "segregation
strength", $\tau $, between positive and negative $\phi $, and the
interaction strength, $\lambda $.
Many years ago by Brazovskii\cite{Brazovskii} argued that this model 
 may undergo a fluctuation
induced first order transition into a lamellar state.

We recently argued that systems described by Eq.\ref{Braz} may also undergo a
self-generated glass transition, i.e. can become non-ergodic without the
presence of quenched disorder\cite{Wu02}. This result was based on our
earlier work\cite{SW,WSW01} which demonstrated such glassiness in a related
model  discussed in the context of frustration limited domain
formation by Kivelson {\em et al.}\cite{Kivelson}. The qualitative features
of that theory were confirmed in mode-coupling calculations and Monte Carlo
simulations of the corresponding lattice model by Grousson {\em et al.}\cite%
{Grousson1,Grousson2}. The theory is based on a replica formalism and two key
assumptions were made in Refs.\onlinecite{Wu02,SW,WSW01}: i) Glassiness can be
studied within the mean-field replica formalism with one step replica
symmetry breaking which was shown to be marginally stable.
 ii) The self-consistent screening approximation (SCSA), including de Gennes narrowing, 
for the summation of the perturbation series in $\lambda $ can be used.
Recently we have been able  to avoid this second assumption and
found a complete solution of the mean field replica equations by generalizing
the dynamic mean field approach (DMFT) of correlated electron models\cite%
{dmft2} to the case of glassy systems.\cite{dmftglass}  Given that our
theory still relies on the applicability of the mean-field replica
formalism, computer simulations like those of Refs.\onlinecite{Grousson1,Geissler}
remain crucial for a final judgment on the existence of self generated
glassiness in uniformly  frustrated systems of this type.

Geissler and Reichman performed Monte-Carlo(MC) calculations and compared
the results with two analytical approaches, one, a Hartree theory and
the other, a mode coupling theory performed self-consistently up to second
order in $\lambda $.  For $\lambda =1$, $k_{0}=0.5$ and $\tau $ between $%
-0.14$ and $0$, good agreement between the MC and Hartree approaches was
obtained, whereas the primitive mode-coupling theory yields glassy dynamics
already for $\tau \leq -0.1$, which was not seen in the Monte Carlo simulations. Thus, the Monte Carlo simulations could not
confirm the glassy dynamics found in mode coupling calculations, but rather
gave only fast, "liquid-like" relaxations. Geissler and Reichman concluded that Eq.\ref{Braz}
likely does not exhibit   glassy dynamics.

\begin{figure}[tbp]
\includegraphics[width=3.0in,angle=-90]{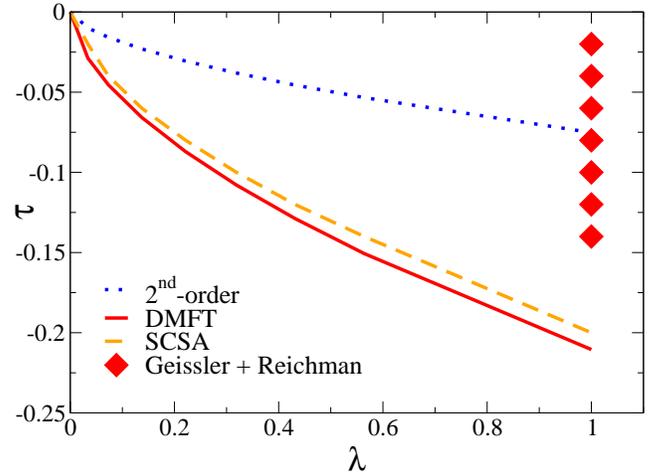}
\caption{Phase segregation strength $\tau$ at the dynamic transition between the
liquid and glass state as function of the coupling constant $\lambda$, obtained by using 2$^{nd}$-order perturbation theory, the self consistent screening approximation(SCSA) of  Refs.\onlinecite{Wu02,SW,WSW01} and the dynamical mean field theory (DMFT) of Ref.\onlinecite{dmftglass}. The diamonds indicate the parameters of the Monte Carlo calculations by Geissler and Reichman where only liquid like behavior was found. The SCSA and DMFT calculations are consistent with the Monte Carlo results whereas the 2$^{nd}$-order calculation (similar to the mode coupling theory  of Ref.\onlinecite{Geissler}) strongly overestimates glassiness.}
\end{figure}

In Fig.1 we compare our results for the dynamic transition between the
liquid and glass state obtained within the replica formalism and solved by
using three different techniques: a self consistent second order theory
(similar to the mode coupling theory of Ref.\onlinecite{Geissler}), the SCSA used
in Refs.\onlinecite{Wu02,SW,WSW01} and the complete solution of the replica mean
field problem (DMFT) of Ref.\onlinecite{dmftglass}. In the second order calculation we determined the
liquid state correlation function within Hartree theory, which seems to be closest to the
 procedure of Ref.\onlinecite{Geissler}, who used Monte Carlo results for the liquid structure facture and find that it is  similar to the one obtained within the Hartree approach. 
The diamonds are
the results of Geissler and Reichman (Figs.5 and 6 of Ref.\onlinecite{Geissler})
where MC simulations give liquid like behavior. It is obvious from this plot
that the results by Geissler and Reichman indeed demonstrate that a second order
mode coupling theory gives a poor approximation for the location of the
glass transition. However, their results cannot be used to rule out the
glass transition we proposed. Within the complete solution of the replica
mean field theory this transition should occur only for $\tau $-values smaller than $%
-0.2$. In fact Fig. 2 of Geissler and Reichman is not inconsistent with a
very slow relaxation rate for $\tau \lesssim -0.2$.  Clearly, just as for
any phase transition, even the best solution of the mean field approach
might yield a phase boundary which is different from what the essentially
exact Monte Carlo simulation finds. The key aspect of our theory was  that
even a simple mean field theory leads to a proliferation of metastable
states and to an entropy crisis. Thus, self generated randomness emerges 
according to the random first order transition scenario.\cite{KTW89a} 

That the parameters studied  in Ref.\onlinecite{Geissler} are outside the region
relevant for glassiness is consistent with their findings that no indication
for the first order Brazovskii transition was detected either.  We argued
in Ref.\onlinecite{Wu02} that the glass transition occurs due to the same type of
fluctuations and for similar parameters to the fluctuation induced first
order transition proposed by Brazovskii\cite{Brazovskii}. The absence of the
glass transition could therefore most convincingly be demonstrated by showing the impossibility of achieving  cooling rates that exceed nucleation rates.

In summary, the recent Monte Carlo calculations by 
Geissler and Reichman\cite{Geissler} are not conclusive in ruling out the self-generated glass
transition proposed  in Ref.\onlinecite{SW}. Nevertheless, if
performed for parameters relevant for glassy behavior Monte Carlo
calculations like theirs would be very useful in judging whether Eq.\ref{Braz} can be
considered as a minimal model for glassiness or whether additional ingredients are needed.
 In addition they offer a
powerful tool to study glassy dynamics beyond the limitations of replica
mean field theory.

This work has been supported by Research Corporation \ and Ames Laboratory,
operated for the U. S. Department of Energy by Iowa State University under
Contract No. W-7405-Eng-82 (J. S. and S.W.) and NSF (P. G. W.), Grant No.
ChE-9530680.

\bigskip \noindent J\"{o}rg Schmalian, Sangwook Wu

{\footnotesize Department of Physics and Astronomy and }

{\footnotesize Ames Laboratory, Iowa State University, Ames, IA 50011}

\noindent Peter G. Wolynes,

{\footnotesize Department of Chemistry and Biochemistry, }

{\footnotesize University of California at San Diego, La Jolla, CA 92093}


\begin{thebibliography}{99}
\bibitem{fn1} In Ref.[1] there is no factor $\frac{1}{2}$ in front of the
term quadratic in $\phi $. Since this factor is needed to derive Eqs. 7 and
9 from Eq.2 of Ref.[1], we assume that the missing pre-factor is a misprint
of Ref.[1].
\end{thebibliography}
\end{document}